\documentclass[12pt]{article}
\usepackage{pdflscape}
\usepackage{amsmath,amssymb,amsthm,amsxtra,overpic,bbm,bm,epsfig,subfigure}
\usepackage{mathrsfs}
\usepackage{booktabs}
\usepackage{graphicx}
\usepackage{xcolor}
\usepackage{comment}
\usepackage{epstopdf}
\usepackage{float}
\usepackage{cite}
\usepackage{array}
\usepackage{makecell}
\usepackage{enumitem}
\usepackage{tabularx} 
\usepackage{overpic}
\usepackage{fancyhdr}
\usepackage{float}
\usepackage{rotating}
\usepackage{color}
\usepackage{esint} 

\allowdisplaybreaks

\makeatletter
\usepackage{slashed,stmaryrd}
\def\thefootnote{\fnsymbol{footnote}}
\newcommand{\myparallel}{{\mkern3mu\vphantom{\perp}\vrule depth 0pt\mkern3mu\vrule depth 0pt\mkern3mu}}
\usepackage{geometry}
\definecolor{MyDarkBlue}{rgb}{0.1, 0.1, 0.8} 
\definecolor{SBlue}{rgb}{0.2, 0.4, 0.7} 
\definecolor{MyLightBlue}{rgb}{0.22,0.51,0.9}
\definecolor{MyGreen}{rgb}{0.0, 0.5, 0.0}
\definecolor{BrickRed}{rgb}{0.8, 0.25, 0.33}
\usepackage{hyperref}
\hypersetup{colorlinks, citecolor=SBlue,linkcolor=MyDarkBlue, urlcolor=BrickRed}

\begin{document}
\begin{center}
{\Large \bf 
Neutrino-antineutrino Asymmetry of C$\bm{\nu}$B on the Surface of the Round Earth
}
\end{center}

\renewcommand{\thefootnote}{\fnsymbol{footnote}}
\vspace{0.05in}
\begin{center}
{
{}~\textbf{Guo-yuan Huang$^{1,2}$}\footnote{ E-mail: \textcolor{MyDarkBlue}{huanggy1992@gmail.com}} 
}
\vspace{0.1cm}
{
\\
\em $^1$School of Mathematics and Physics, China University of Geosciences, 430074 Wuhan, China\\
$^2$Max-Planck-Institut f{\"u}r Kernphysik, Saupfercheckweg 1, 69117 Heidelberg, Germany
} 
\end{center}

\renewcommand{\thefootnote}{\arabic{footnote}}
\setcounter{footnote}{0}
\thispagestyle{empty}
\vspace{0.5cm}
\begin{abstract}
\noindent It has been claimed that the coherent scattering of relic neutrinos with the Earth will result in a neutrino-antineutrino asymmetry of $\mathcal{O}(10^{-4})$ on the Earth surface, which is five orders of magnitude larger than the naive model expectation. In this work we show that this overdensity was overestimated for the perfectly round Earth by solving the exact solution with partial waves. The maximal asymmetry after summing over all the angular modes is only  around $10^{-8}$ above the ground. To achieve the proposed asymmetry of $\mathcal{O}(10^{-4})$, a special geography may be needed as the experimental site.
\end{abstract}
\setcounter{footnote}{0}

\newpage

\newpage 

\section{Introduction}
The detection of relic neutrinos remains one of the most fundamental challenges of experimental particle physics. 
The most promising way so far is by observing the inverse beta decays, proposed by S.~Weinberg in 1962~\cite{Weinberg:1962zza}, even though there are still several technical obstacles towards a feasible observation~\cite{PTOLEMY:2018jst,PTOLEMY:2019hkd,PTOLEMY:2022ldz,Cheipesh:2021fmg,Cheipesh:2023qiy}.
Besides this method, an alternative is to observe the collective effects induced by the wind of relic neutrinos~\cite{Opher:1974drq,Stodolsky:1974aq,Lewis:1979mu,Zeldovich:1981wf,Cabibbo:1982bb,Shvartsman:1982sn,Langacker:1982ih,Smith:1983jj,Lewis:1987yd,Loeb:1990xs,Ferreras:1995wf,Hagmann:1999kf,Duda:2001hd,Gelmini:2004hg,Ringwald:2009bg,Vogel:2015vfa,Domcke:2017aqj,Shergold:2021evs,Arvanitaki:2022oby,Ruzi:2023cvp,Arvanitaki:2023fij}.
Compared to the radioactive process, the major advantage of coherent scatterings is that the event rate can be enhanced by a factor as large as the Avogadro constant $N^{}_{\rm A} \approx 6.02 \times 10^{23}$, which may help to compensate the suppression of Fermi coupling constant $G^{}_{\rm F}$.
There are mainly two effects that have been investigated in this direction: one is the mechanical recoil induced by the net collision; the other is the Stodolsky effect~\cite{Stodolsky:1974aq}, i.e., a net torque due to the spin energy splitting in the presence of a neutrino-antineutrino asymmetry.
The challenge of the first effect is that the mechanical recoil will be distributed among all the atoms in the ensemble, and hence the ultimate induced acceleration is highly suppressed.
Whereas, the Stodolsky effect is proportional to the neutrino-antineutrino asymmetry $\eta^{}_{\nu\overline{\nu}}$, which is expected to be of the same order of magnitude as the Baryon asymmetry $\mathcal{O}(10^{-9})$ from many models.
This number is too small to have any observable effects.

Surprisingly, it has been recently noticed that this asymmetry can be enhanced to $\eta^{}_{\nu\overline{\nu}} = \mathcal{O}(10^{-4})$ due to the reflection effect on the Earth surface~\cite{Arvanitaki:2022oby}.
This is not difficult to understand as neutrinos and antineutrinos feel opposite matter potentials of the Earth.
In particular, the total external reflection occurs for neutrinos as in optics when the incidence angle of relic neutrino flux approaches the critical angle.
With an isotropic neutrino flux, the reflected waves were claimed to result in a large neutrino-antineutrino asymmetry near the Earth surface, which is five orders of magnitude larger than $\mathcal{O}(10^{-9})$.
If confirmed, this effect would be very encouraging for the observation of relic neutrinos.

The above conclusion was primarily drawn by approximating the Earth to the toy model of a slab with an infinite thickness, and neutrinos were assumed to be injected from vacuum above the slab surface.
For the round Earth with a finite size, this approximation appears to be oversimplified, because   trajectories of any incident neutrinos will eventually emerge from the Earth surface. A better approximation of the round Earth might be a slab with a finite thickness.
The thickness of the slab can be roughly represented by the depth that neutrinos can reach inside the Earth with a certain incidence angle.
The neutrino overdensity on the Earth surface is in principle given by the neutrino wavefunction integrated over all incoming angles.
In this case, there will be neutrinos injected from both above and below one slab surface.
Unfortunately, we find that accounting for those neutrinos coming from the other side of the slab will reduce the overdensity significantly.
To reinforce our result, we have calculated this overdensity without the slab approximation, by directly considering the potential of the round Earth and solving the scattering problem with partial waves.

The structure of the rest of this work is as follows. In Sec.~\ref{sec:II}, we investigate the ideal reflection of relic neutrinos by a finite-thickness slab, and discuss how the overdensity will be altered for an isotropic neutrino injection. In Sec.~\ref{sec:III}, the neutrino wavefunction by collectively scattering with the round Earth will be solved by summing up contributions of different partial waves. We finally make our conclusion and remark on a possible solution in Sec.~\ref{sec:IV}.

\section{Reflection of Neutrinos by a Finite-thickness Slab} \label{sec:II}
\begin{figure}[t!]
	\vspace{-0.1cm}
	\begin{center}
		\includegraphics[width=0.65\textwidth]{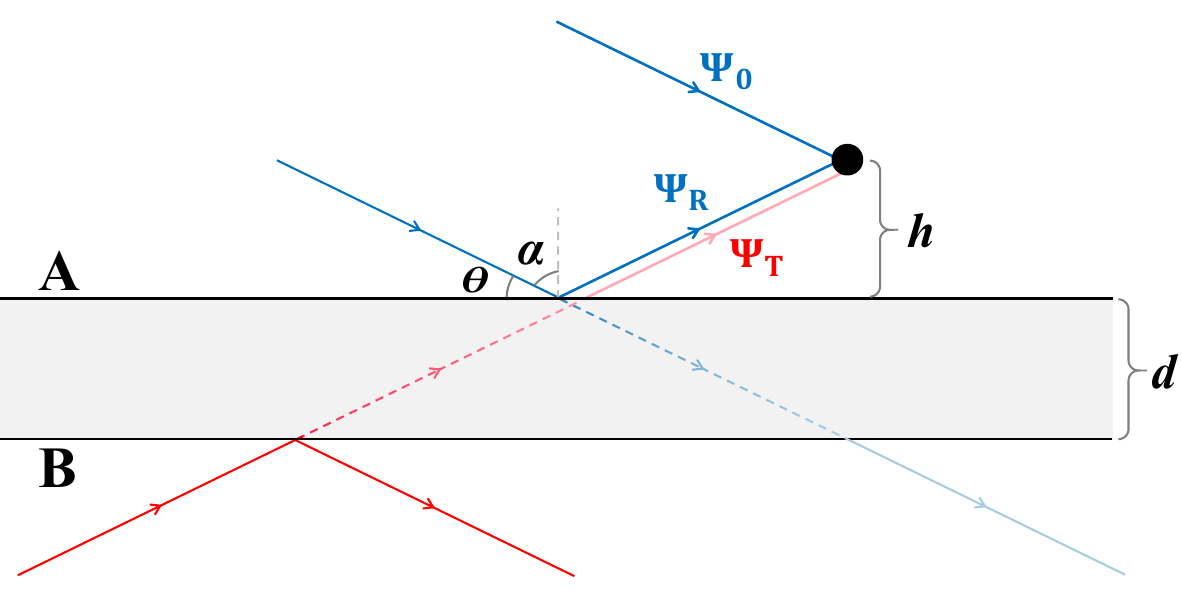}
	\end{center}
	\vspace{-0.5cm}
	\caption{A schematic plot for the coherent scattering of neutrinos with a slab with the thickness $d$. The detector is placed at height $h$ to the slab surface. The overdensity receives three contributions: (i) the reflected wave solely $|\Psi^{}_{\rm R}|^2$; (ii) the interference between the original and reflected waves $2\mathrm{Re}(\Psi^{}_{\rm 0} \Psi^{*}_{\rm R})$; (iii) the transmitted wave from the other side of the slab $|\Psi^{}_{\rm T}|^2$. }
	\label{fig:slab}
\end{figure}

To solve the neutrino wavefunction, one has to investigate the scattering of incident neutrino waves with a spherical potential.
However, if the effect of curvature of the Earth surface is small enough, one may approximate the scenario of concern to the coherent scattering of plane waves with a finite-thickness slab. 
The depth of the slab will be around $d = R^{}_{} (1-\cos{\theta}) \sim R\, \theta^2/2$ with $\theta$ being the incidence angle with respect to the horizon and $R = 6371~{\rm km}$ being the Earth radius.
The corresponding zenith angle is just $\alpha = \pi/2 - \theta$.
A schematic diagram of the scenario of concern is shown in Fig.~\ref{fig:slab}.

Considering the above approximation, we first investigate the case that a collimated monoenergetic neutrino beam is incident on an infinitely wide slab with the thickness $d$. 
By collectively interacting with the matter in the slab, neutrino waves will undergo both reflection and transmission near the slab surface.
In particular, a positive potential in matter can result in the total external reflection of neutrinos when the incidence angle is smaller than the critical angle $\theta^{}_{\rm c}$.
The detailed outcome of the scattering depends on the neutrino mass, the neutrino momentum and the matter potential. For the convenience of later discussions, we set the neutrino mass to be $m^{}_{\nu} = 0.1~{\rm eV}$ with a momentum $k^{}_{\nu} \approx 2.7 \times 10^{-4}~{\rm eV}$ following Ref.~\cite{Arvanitaki:2022oby}. 
The matter potential describing the coherent forward scattering of electron neutrinos is also chosen to be
$
V_{\nu_{e}}  = - V_{\overline{\nu}_{e}} \approx 1.8 \times 10^{-14}~{\rm eV}, 
$
contributed by both the charged-current and neutral-current interactions.
A convenient dimensionless parameter that quantifies the strength of the matter potential is  defined by $\delta \equiv m^{}_{\nu} V / k^2_{\nu} \approx \pm 2.5 \times 10^{-8}$.
Note that we can safely ignore the recoil of the slab induced by coherent scatterings considering the heaviness of the Earth.

Let us denote the normal vector of the slab surface to be the $z$-axis.
Since this system features a translational symmetry in the $x$-$y$ plane along the slab surface, one can simplify the problem to a one-dimensional scattering with a potential barrier of the thickness $d$. The scattering behavior will hence rely on the relation between the $z$ component of the neutrino momentum $k^{}_{\nu \perp}$ and the potential $V$.
If $k^{}_{\nu \perp}< V$, the transmission will be exponentially suppressed but not entirely vanishing because of the quantum tunneling effect through the barrier.
The reflected and transmitted waves in vacuum can be parameterized by $\Psi^{}_{\rm R} = C^{}_{\rm R} \exp{\mathrm{i} (k^{}_{\nu \myparallel} x - k^{}_{\nu \perp} z)}$ and $\Psi^{}_{\rm T} = C^{}_{\rm T} \exp{\mathrm{i} (k^{}_{\nu \myparallel} x + k^{}_{\nu \perp} z)}$, respectively. 
Inside the matter, the wavefunction is $\Psi^{\prime}_{} = C^{\prime}_{1} \exp{\mathrm{i} (k^{}_{\nu \myparallel} x - k^{\prime}_{\nu \perp} z)} +C^{\prime}_{2} \exp{\mathrm{i} (k^{}_{\nu \myparallel} x + k^{\prime}_{\nu \perp} z)}$ with $k^{\prime}_{\nu \perp} = \sqrt{k^{2}_{\nu \perp} - 2 m^{}_{\nu} V}$ being the effective momentum in matter. Depending on the value of $V$, $k^{\prime}_{\nu \perp}$ can be imaginary which indicates the exponential suppression of wavefunctions in the barrier. The critical angle for total reflection is just $\sin \theta^{}_{\rm c} = \sqrt{2 m^{}_{\nu} V}/k^{}_{\nu}$ set by $k^{\prime}_{\nu \perp} = 0$. For the above chosen parameters, the critical angle is $\sin \theta^{}_{\rm c} \approx 2 \times 10^{-4}$.
One can solve those coefficients by matching the wavefunctions and their derivatives at the boundaries of the slab.
We find the following solutions to the reflection and transmission coefficients,
\begin{align}
C^{}_{\rm R} & =  \frac{(k^{2}_{\nu \perp} - k^{\prime 2}_{\nu \perp})  \sin(d k^{\prime}_{\nu \perp} )}{ (k^{2}_{\nu \perp} + k^{\prime 2}_{\nu \perp})  \sin(d k^{\prime}_{\nu \perp} ) +2\,\mathrm{i}\, k^{}_{\nu \perp} k^{\prime}_{\nu \perp} \cos(d k^{\prime}_{\nu \perp} ) }\;, \\
C^{}_{\rm T} & =  \frac{2\, \mathrm{i} \, \mathrm{e}^{-\mathrm{i} d k^{}_{\nu \perp}} k^{}_{\nu \perp} k^{\prime}_{\nu \perp}}{ (k^{2}_{\nu \perp} + k^{\prime 2}_{\nu \perp})  \sin(d k^{\prime}_{\nu \perp} ) +2\,\mathrm{i}\, k^{}_{\nu \perp} k^{\prime}_{\nu \perp} \cos(d k^{\prime}_{\nu \perp} ) }\;.
\end{align}
The above expressions apply to all values of $V$.  For $k^{}_{\nu \perp}< V$, one may find it convenient to replace $k^{\prime}_{\nu \perp}$ by $\mathrm{i} \gamma$ with $\gamma = \sqrt{2 m^{}_{\nu} V - k^{2}_{\nu \perp}}$.
In the limit of infinity thickness, we have
\begin{align}
	C^{}_{\rm R} & =  \frac{k^{}_{\nu \perp}-k^{\prime}_{\nu \perp}}{k^{}_{\nu \perp}+k^{\prime}_{\nu \perp}}\;.
\end{align}
For the infinity thickness, the total reflection occurs in the case of imaginary $k^{\prime}_{\nu \perp}$, i.e., $|C^{}_{\rm R}|^2 = 1$. 
However, we should emphasize that this is not true anymore for the finite thickness.
In particular, for small $d$, the reflection probability takes a perturbative value in the Born approximation~\cite{Smith:1983jj}
\begin{align}
	|C^{}_{\rm R}|^2 & \approx  \frac{( m^{}_{\nu} V)^2 d^2}{ k^{2}_{\nu \perp}}\;.
\end{align}
This probability is also suppressed by the smallness of the matter potential.
Around the critical angle, the condition for $|C^{}_{\rm R}|^2$ to be order one is given by $d > (m^{}_{\nu} V)^{-1/2} \sim 5~{\rm m}$.
In comparison, the depth that neutrinos can travel inside the Earth at the critical angle $\theta^{}_{\rm c} \sim 2.2 \times 10^{-4}$ is only $0.16~{\rm m}$, and correspondingly $|C^{}_{\rm R}|^2 \sim 6 \times 10^{-4}$, which is still within the valid regime of the Born approximation.
This is well understood as sufficient matter  is required along the neutrino path for the non-perturbative total reflection to happen.
As a result, the total reflection over a flat slab may not be fully applicable to our round Earth.
Nevertheless, it is worthwhile to first analyze the common features by exploring the overdensity in the simplified scenario with a sufficiently large thickness.
We will solve the overdensity without such approximations in the next section.

\begin{figure}[t!]
	\vspace{-0.1cm}
	\begin{center}
		\includegraphics[width=0.45\textwidth]{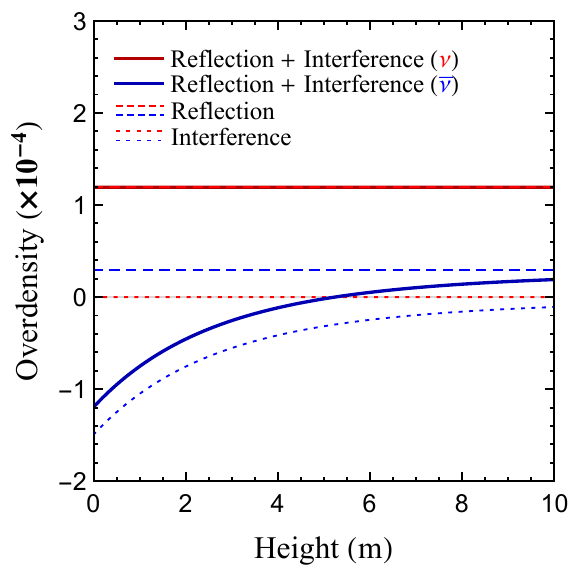}
	\end{center}
	\vspace{-0.5cm}
	\caption{The neutrino (in red) and antineutrino (in blue) overdensities in terms of the height $h$ to the slab surface. Separate contributions are shown from the reflected wave alone (dashed curves) and from the interference between the original and reflected waves (dotted curves). }
	\label{fig:dnnu_slab}
\end{figure}

Above the slab surface, the neutrino overdensity is contributed by the reflected wave as well as the interference between the original and reflected waves.
In some cases, the interference term can even be dominant.
The total overdensity should be obtained by integrating over the zenith angle $\alpha$ of incoming neutrinos. Following Ref.~\cite{Arvanitaki:2022oby}, if we account for those neutrinos only from above the slab, the overdensity reads
\begin{align} \label{eq:dnnu_slab}
\delta n^{}_{\nu(\overline{\nu})} = \int^{1}_{0} \left[\frac{|\Psi^{}_{\rm R}|^2}{2} + \mathrm{Re}(\Psi^{}_{\rm 0} \Psi^{*}_{\rm R}) \right] \mathrm{d} \cos{\alpha} \;,
\end{align}
The asymmetry is then given by $\eta^{}_{\nu\overline{\nu}} = (\delta n^{}_{\nu} - \delta n^{}_{\overline{\nu}}) / n^{}_{\nu}$.
In Fig.~\ref{fig:dnnu_slab}, we plot the overdensities $\delta n^{}_{\nu}$ and $\delta n^{}_{\overline{\nu}}$ in terms of the height to the slab surface $h$.
To analyze the importance of different terms, we have also depicted separate contributions from the reflected wave $|\Psi^{}_{\rm R}|^2$ and the interference $2\mathrm{Re}(\Psi^{}_{\rm 0} \Psi^{*}_{\rm R})$.
The total overdensities for neutrinos and antineutrinos are shown as the solid red and blue curves, respectively, whose difference can reproduce the asymmetry in Ref.~\cite{Arvanitaki:2022oby}.
For neutrinos, the overdensity of $\mathcal{O}(10^{-4})$ is dominated by the reflected wave solely (dashed red curve), while the interference contribution (dotted red curve) is negligible. For antineutrinos, both the reflection and the interference are important in general.

As has been mentioned, a loophole of the above analysis is that neutrinos coming from the other side of the slab are not taken into account. We find that accounting for those flux will in general reduce the overdensity of neutrinos.
In particular, we can observe the following fact: the overdensity induced by the total reflection over one side of the slab is right equal to the deficit of transmitted waves from the other side due to the probability conservation, i.e., $|\Psi^{}_{\rm R}|^2 = 1 - |\Psi^{}_{\rm T}|^2$.
Hence, only the negligible interference contribution remains if we add them up.
Nevertheless, the above argument cannot be simply applied to antineutrinos, for which the interference might be important. For antineutrinos with small heights in Fig.~\ref{fig:dnnu_slab}, the deficit from the other side will make the total overdensity even more sizable.
However, this is not true for the round Earth in practice. A careful investigation shows that the round Earth with a negative potential serves as a convex lens. For the parameter space relevant for the Earth, the interference between different  paths of transmitted waves will also compensate for the original negative overdensity of antineutrinos.
In general, the overdensities for both neutrinos and antineutrinos (and hence the asymmetry) are expected to be severely suppressed compared to the case with flux injection only from one side. However, the exact value can only be obtained by solving the scattering with the spherical potential of the Earth.

\section{Suppressed Asymmetry for Perfectly Round Earth} \label{sec:III}

\begin{figure}[t!]
	\vspace{-0.1cm}
	\begin{center}
		\includegraphics[width=0.8\textwidth]{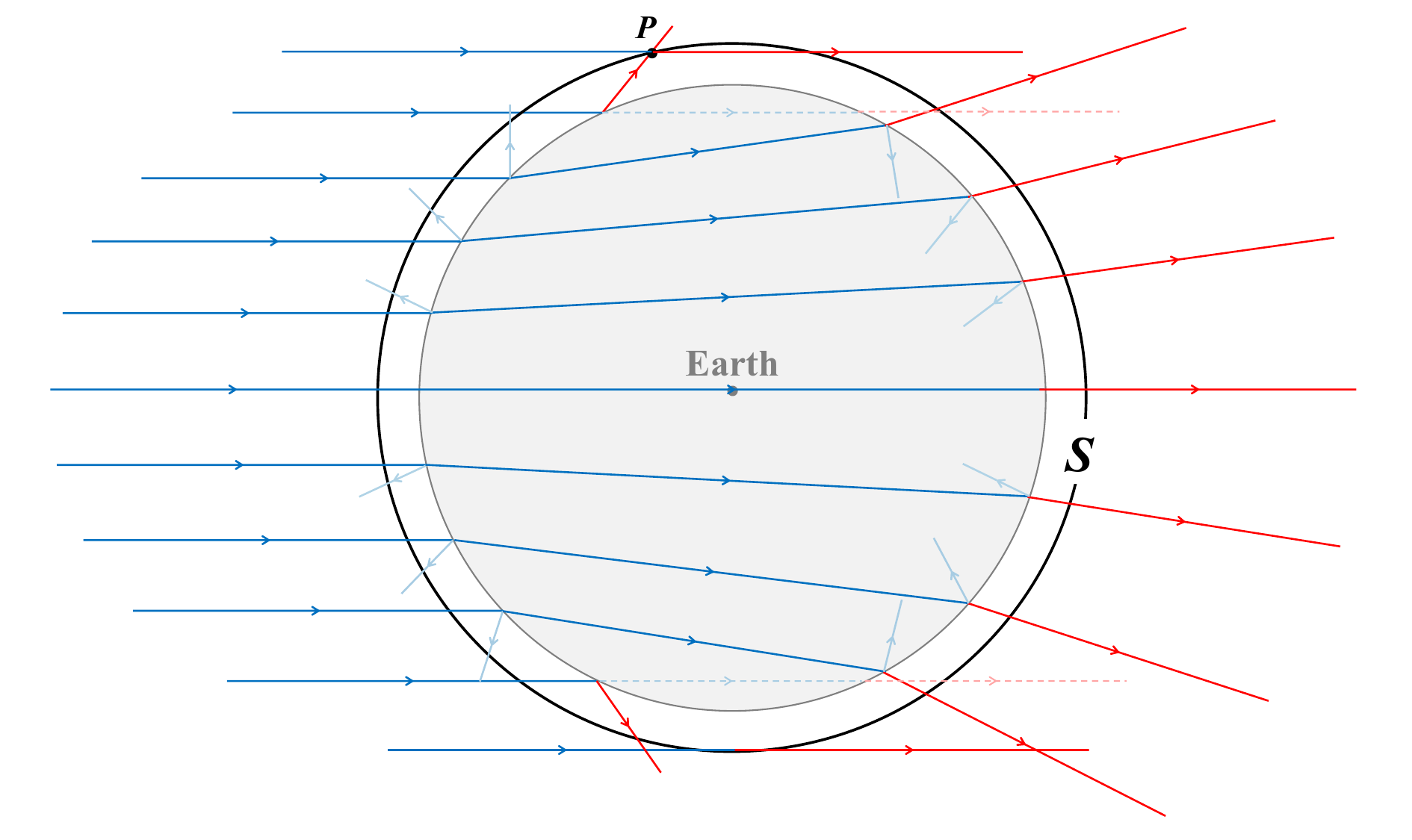}
		\includegraphics[width=0.8\textwidth]{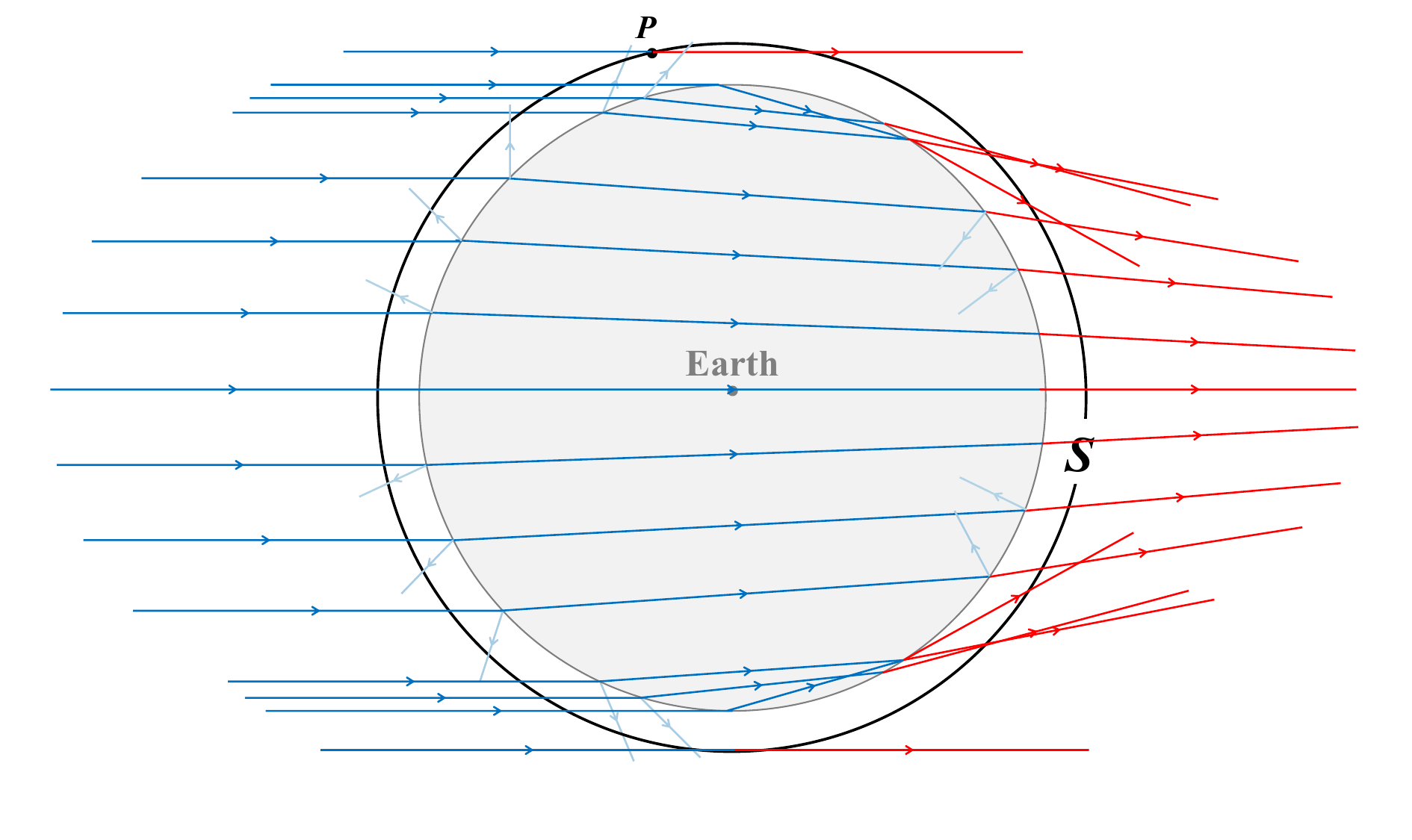}
	\end{center}
	\vspace{-0.5cm}
	\caption{An illustration of a plane-wave neutrino (upper panel) or antineutrino (lower panel) scattering with the matter potential of the Earth. The detector is placed at the height of $P$. The neutrino density at $P$ with an isotropic flux injection is equivalent to integrating the wavefunction over the sphere $S$ with a collimated flux. }
	\label{fig:Earth}
\end{figure}

In this section, we prove that the neutrino-antineutrino asymmetry is highly suppressed for the perfectly round Earth with an isotropic injection of the cosmic neutrino flux. 
This result is valid because both the flux input and the matter potential under assumption feature the spherical symmetry. 
We shall calculate the neutrino number density $n^{}_{\nu}$ at a given point $P$ above the Earth. 
For illustration, a schematic diagram for the scattering of neutrinos with a positive potential of the Earth is given in Fig.~\ref{fig:Earth}.

As the first step, we notice that the spherical symmetry enables us to translate the neutrino number density at point $P$ with the isotropic flux injection into an integration of the density over a sphere $S$ containing $P$ but with a {\it collimated} injection. 
The density value at each point on the sphere is identical to the density at the fixed point $P$ contributed by the flux from a different incoming angle (rotate that point to match with $P$).
This equivalence is critical for our later understanding from the classical perspective. 

Suppose that a monoenergetic plane wave is incident on the Earth with the free-particle wavefunction $\phi(\bm{x}) = \mathrm{exp}(\mathrm{i} \bm{k} \cdot \bm{x})$ far away from the potential. The resultant wavefunction by scattering with the Earth is denoted as $\Psi(\bm{x})$. In the absence of the Earth, we should have $\Psi(\bm{x}) = \phi(\bm{x})$.
The number density with an isotropic flux injection hence reads
\begin{align} \label{eq:nnu}
	n^{}_{\nu (\overline{\nu})}(r) =  \frac{n^{}_{0}}{4\pi} \int \left| \Psi(\bm{x}) \right|^2 \mathrm{d}\Omega = \frac{n^{}_{0}}{4\pi r^2} \varoiint  \left| \Psi(\bm{x}) \right|^2 \mathrm{d}S \;,
\end{align}
where the normalization factor $n^{}_{0}$ is just the number density of neutrinos in the absence of the Earth, $r$ is the distance from the Earth center to the detector, and the second integration is performed over the sphere $S$. 

\subsection{The Semiclassical Approximation}
We attempt to prove that $n^{}_{\nu}(r) = n^{}_{0}$ for both neutrinos and antineutrinos (no overdensity for the round Earth) in the semiclassical limit by considering neutrinos as localized particles following certain trajectories.
We will find that the result holds because of several conservation laws.
In the semiclassical limit, the interference effect between different waves has been ignored.
It is then straightforward to validate the result by explicitly counting the neutrino flux going into (in blue) and out of (in red) the sphere $S$ as shown in Fig.~\ref{fig:Earth}.
From the particle number conservation, we may first establish the following relation 
\begin{align}
n^{}_{\rm in} \bm{v}^{}_{\rm in} \cdot \mathrm{d} \bm{S} = n^{}_{\rm out} \bm{v}^{}_{\rm out} \cdot \mathrm{d} \bm{S}^{\prime} \;,
\end{align}
where $n^{}_{\rm in (out)}$ is the number density of the neutrino flux going into (out of) the sphere $S$, the flux strength is just given by $n \bm{v}$ with $\bm{v}$ being the corresponding velocity, and $\mathrm{d} \bm{S}^{\prime}$ is a small area on the sphere which the neutrino flux from $\mathrm{d} \bm{S}$ is deflected into (e.g., by total reflection).

Let us denote the angles between the neutrino velocity and the normal vector of the area as $\theta^{}_{\rm in}$ and $\theta^{}_{\rm out}$, respectively.
According to the energy conservation, ${v}^{}_{\rm in} = {v}^{}_{\rm out}$ must hold. 
Furthermore, the angular momentum conservation should be satisfied for the spherical potential due to rotational symmetry, leading to $\theta^{}_{\rm in} = \theta^{}_{\rm out}$. 
As a consequence, we obtain
$
n^{}_{\rm in} \cdot \mathrm{d} {S} = n^{}_{\rm out} \cdot\mathrm{d} {S}^{\prime}
$. The number density with the isotropic injection is then
\begin{align}
	n^{}_{\nu (\overline{\nu})}(r) = \frac{1}{4\pi r^2} \varoiint (n^{}_{\rm in} + n^{}_{\rm out}) \mathrm{d}S = \frac{2}{4\pi r^2} \varoiint n^{}_{\rm in} \mathrm{d}S \;.
\end{align}
Here, $n^{}_{\rm in} = n^{}_{0}/2$, which is independent of how neutrinos are deflected by the potential in the Earth.
Thus, we have $n^{}_{\nu}(r) = n^{}_{0}$ in the semiclassical limit.

However, the above conclusion is meaningful only when the semiclassical viewpoint of particles with trajectories is involved.
In practice, there will also be diffraction and interference effects between the incoming and outgoing waves.
To account for those contributions, we attempt to solve the neutrino wavefunction exactly.

\subsection{A Rigorous Treatment}
The scattering problem of concern is very typical and can be solved following the textbook~\cite{Sakurai:2011zz}.
The task is to find the solution of the Lippmann-Schwinger equation
\begin{align} \label{eq:LSE}
\left| \Psi \right\rangle = \left| \phi \right\rangle + \frac{1}{E-\widehat{H}^{}_{0} + \mathrm{i} \epsilon } \widehat{V}  \left| \Psi \right\rangle \;,
\end{align}
where $\phi$ is the free wavefunction, $\widehat{H}^{}_{0}$ is the free Hamiltonian, and $\widehat{V}$ is the potential.
Because the Earth potential is taken to be spherically symmetric, a general form of the neutrino wavefunction outside the Earth can be found with the help of partial-wave expansion,
\begin{align}
	\Psi(r,\theta) = \sum^{}_{l} \mathrm{i}^l (2l+1) A^{}_{l}(r) P^{}_{l}(\cos{\theta}) \;,
\end{align}
where $\theta$ is the zenith angle with the $z$-axis fixed by the direction of asymptotic incoming wave, $P^{}_{l}$ is the Legendre polynomial, and $l$ represents the angular momentum. 
Adopting Eq.~(\ref{eq:LSE}), the radial wavefunction $A^{}_{l}(r)$ in the non-relativistic limit follows
\begin{align}
\frac{\mathrm{d}^2 (r A^{}_{l})}{\mathrm{d} r^2} + \left[k^2_{\nu} - 2 m V(r) - \frac{l(l+1)}{r^2} \right] r A^{}_{l} = 0 \;,
\end{align}
where $V(r) = V \Theta(R-r)$ with $R$ being the Earth radius. 
The solution can be conveniently expressed in terms of the Bessel functions
\begin{align}
	A^{}_{l} (r) = \mathrm{e}^{\mathrm{i} \delta^{}_{l}} \left[ \cos{\delta^{}_{l}} j^{}_{l}(k^{}_{\nu} r) - \sin{\delta^{}_{l}} n^{}_{l}(k^{}_{\nu} r) \right] ,
\end{align}
where $j^{}_{l}$ and $n^{}_{l}$ are the spherical Bessel functions of first and second kinds, respectively, and $\delta^{}_{l}$ is the phase shift determined by the spherical potential.
Similar to the case of the slab, by matching the wavefunction at boundaries, we obtain the phase shift
\begin{align}
\tan{\delta^{}_{l}} = \frac{k^{}_{\nu}\, j^{}_{l}(k^{\prime}_{\nu} R) \, j^{}_{l+1}(k^{}_{\nu} R) - k^{\prime}_{\nu}\, j^{}_{l}(k R)\, j^{}_{l+1} (k^{\prime}_{\nu} R) }{k^{}_{\nu}\, j^{}_{l}(k^{\prime}_{\nu} R)\, n^{}_{l+1}(k^{}_{\nu} R) - k^{\prime}_{\nu}\, j^{}_{l+1}(k^{\prime}_{\nu} R)\, n^{}_{l}(k^{}_{\nu} R) } \;,
\end{align}
with $k^{\prime}_{\nu} = (k^2_\nu - 2m V)^{1/2}$ being the effective momentum in matter. The above phase shift will become vanishingly small around $l \gtrsim k^{}_{\nu} R \approx 8.7 \times 10^9$, which makes sense as angular modes with impact parameters larger than the Earth radius should not contribute to the scattering.

Using the above representation, the neutrino number density obtained with Eq.~(\ref{eq:nnu}) for the isotropic injection turns out to be
\begin{align}
	n^{}_{\nu (\overline{\nu})}(r) = \frac{n^{}_{0}}{2} \int^1_{-1} \left| \Psi(r,\theta) \right|^2 \mathrm{d} \cos\theta = n^{}_{0} \sum^{}_{l} (2l+1) |A^{}_{l}(r)|^2\;,
\end{align}
In deriving the above relation, the orthogonal relation of different angular modes has been used.
We can check the validity of the above relation in the absence of the Earth potential, for which the phase shift is vanishing, i.e., $\delta^{}_{l} = 0$. This is easy to verify by noticing the sum rule $\sum^{\infty}_{l = 0} (2l+1) j^2_{l}(kr) = 1$.
For non-vanishing phase shifts, we find the overdensity $\delta n^{}_{\nu (\overline{\nu})}(r) \equiv n^{}_{\nu (\overline{\nu})}(r) - n^{}_{0}$,
\begin{align} \label{eq:dnnur}
	\delta n^{}_{\nu (\overline{\nu})}(r) = n^{}_{0} \sum^{\infty}_{l = 0} (2l+1) \left[\sin^2\delta^{}_{l} \, (n^{2}_{l}(k r)-j^{2}_{l}(k r))- \sin 2\delta^{}_{l}\, j^{}_{l}(k r) n^{}_{l}(kr)  \right] .
\end{align}
The total reflection is supposed to occur at very small incidence angles, corresponding to $l \sim k R$, and hence the above summation should be dominated by those terms around $l \lesssim k R$. However,
we can first check the asymptotic behavior of Bessel functions.
In the limit of $kr \to \infty$ with $l$ being small, the spherical Bessel functions can be approximated as
\begin{align} \label{eq:jlkr}
	j^{}_{l} (kr) \approx \frac{1}{kr} \cos\left[kr - \frac{(l-1) \pi}{2} \right] ,\;\;\; n^{}_{l} (kr) \approx \frac{1}{kr} \sin\left[kr - \frac{(l-1) \pi}{2} \right] .
\end{align}
Those two terms in Eq.~(\ref{eq:dnnur}) are then proportional to $\cos(2 kr - l\pi)$ and $\sin(2 kr - l\pi)$, respectively. Because of the largeness of $r$, those rapidly oscillating terms will be easily averaged to zero in the presence of a tiny dispersion in $k$.
Furthermore, even if $k$ is strictly monoenergetic, contributions of adjacent $l$ in Eq.~(\ref{eq:dnnur}) are found to cancel each other, e.g., $\sin(2 kr - l\pi) + \sin(2 kr - (l+1)\pi) = 0$.

\begin{figure}[t!]
	\vspace{-0.1cm}
	\begin{center}
		\includegraphics[width=0.6\textwidth]{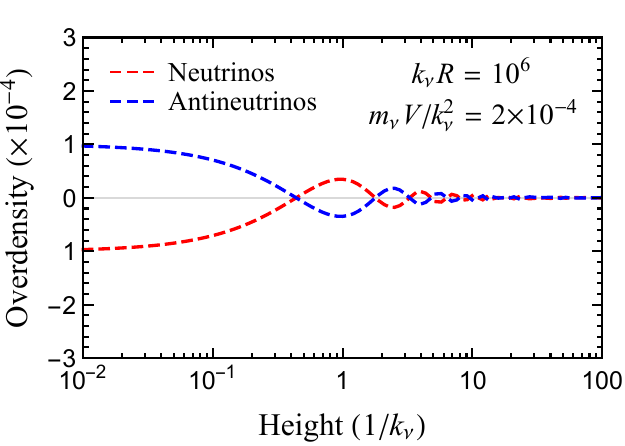}
	\end{center}
	\vspace{-0.5cm}
	\caption{The overdensities of neutrinos (red curve) and antineutrinos (blue curve)  in terms of the height $h$ for a toy parameter choice $k^{}_{\nu} R = 10^6$ and  $ m^{}_{\nu} V /{k^2_{\nu}} = 2 \times 10^{-4}$. For the Earth with $k^{}_{\nu} R \approx 8.73 \times 10^9$, the magnitude of overdensity on the ground is roughly $1.2 \times 10^{-8}$. }
	\label{fig:dnnu_Earth}
\end{figure}

In application, we have to directly sum over all the partial waves without the above approximation. 
To increase the evaluation speed, we have adopted the asymptotic expansions of Bessel functions with large orders of $l$~\cite{NIST:DLMF}. 
Because the effective scattering is largely within the perturbative regime, the contributions from large and small $l$'s are both very important.
By summing over all the angular modes from $l=0$ to $l \gtrsim k^{}_{\nu} R \approx 8.73 \times 10^9$, we obtain the overdensities  on the ground  
\begin{align}
	\delta n^{}_{\nu} \approx -1.21 \times 10^{-8}~\text{and}~\delta n^{}_{\overline{\nu}} \approx 1.26 \times 10^{-8} \;,
	\end{align}
which are roughly of the order of $ m^{}_{\nu} V /{(2 k^2_{\nu})}$.
Away from the ground, the decoherence effect should even reduce the overdensity.
However, because the computation is rather time-consuming, we are unable to finish the scanning over the detector height in a reasonable time.
Instead,  to demonstrate the decoherence effect we choose a relatively small toy model compared to the Earth  with $k^{}_{\nu} R = 10^6$ and  $ m^{}_{\nu} V /{k^2_{\nu}} = 2 \times 10^{-4}$.
In Fig.~\ref{fig:dnnu_Earth}, we show the resultant overdensities as functions of the height $h$ (in units of $1/k^{}_{\nu}$) for both neutrinos (red curve) and antineutrinos (blue curve).
For small heights, the overall overdensity is of $ m^{}_{\nu} V /{(2 k^2_{\nu})}$, similar to the findings for the case of the Earth. For $h > 1/{k^{}_{\nu}}$, the overdensity quickly decreases with respect to the height.
As we mentioned, this overdensity arises from the interference between different waves. Away from the ground, the interference is expected to be vanishingly small for the round Earth.  

\section{Concluding Remarks} \label{sec:IV}
The feasibility of observing relic neutrinos via collective effects will be greatly enhanced,
if there is a large neutrino-antineutrino asymmetry on the surface of the Earth.
However, we have shown that the overdensities of both neutrinos and antineutrinos should be suppressed due to several conservation laws including those of the particle number, the energy and the angular momentum.
The residual factor that can still affect the overdensity is the interference effect between different waves. 
By solving the scattering of plane-wave neutrinos with the spherical potential of the Earth, we have obtained the overdensity on the Earth surface without the flat slab approximation. The overall magnitude at small heights is only of $\mathcal{O}(10^{-8})$, and this value will become vanishingly small away from the ground.

Nevertheless, we should emphasize that the conclusion holds for a perfectly round Earth.
Thus, one possible way out is to choose a patch of ground disobeying the rotational symmetry.
On the one hand, to appreciate the total reflection one should have a rather smooth ground in sight (of area $2 h R$ within the horizon). For $h = 10~{\rm m}$, the dimension of this ground is around $11~{\rm km}$. One the other hand, to deviate from the rotational symmetry one can require the ground beyond the horizon to be sufficiently bumpy. In Fig.~\ref{fig:slab}, this requirement is equivalent to making side B of the slab bumpy such that the total reflection does not occur on that side.
In this manner, one may have a sizable overdensity on the surface of side A without cancellations. \\

\noindent {\bf Note added:} The preprint Ref.~\cite{Gruzinov:2024ciz}, which just appeared on arXiv, has explored the same subject with the thermal method instead of directly summing over the partial waves. Their analysis gives an overdensity of the order of $\mathcal{O}(10^{-8})$, which is in agreement with the findings in this work after summing over all the angular modes.

\section*{Acknowledgments}
I would like to thank Andreas Trautner and Manibrata Sen for insightful discussions during the refinement of this manuscript.
I am also grateful to Prof. Asimina Arvanitaki for valuable communication and for the sharp observation of the overdensity results from some toy computations.

\appendix

\bibliographystyle{utcaps_mod}

\bibliography{reference}

\end{document}